\documentclass[11pt]{article}
\usepackage{graphics,epsfig,subfigure}
\usepackage{color}
\usepackage{diagbox}
\usepackage{amsthm}
\usepackage{amscd}
\usepackage{amstext}
\usepackage{soul} %delet this package after check.
\newcommand{\Appendix}[1]{
\refstepcounter{section}
\makeatletter
\newcommand{\rmnum}[1]{\romannumeral #1}
\newcommand{\Rmnum}[1]{\expandafter\@slowromancap\romannumeral #1@}
\renewcommand{\thefootnote}{}
\footnotetext{Footnotetext without footnote mark}
\makeatother
\begin{flushleft}
{\large\bf Appendix \thesection: #1}
\end{flushleft}}

\newcommand{\ba}{\begin{array}}
\newcommand{\ea}{\end{array}}

\def\be{\begin{equation}}
\def\ee{\end{equation}}
\def\ba{\begin{array}}
\def\ea{\end{array}}

\def\dalemb#1#2{{\vbox{\hrule height .#2pt
        \hbox{\vrule width.#2pt height#1pt \kern#1pt
                \vrule width.#2pt}
        \hrule height.#2pt}}}

\def\ocal{{\mathcal{O}}}

\begin{document}
\begin{center}
%\vspace{1cm} { \LARGE {\bf Holographic excited  superconductors}}

%\vspace{1cm} { \LARGE {\bf Excited holographic superconductors}}

\vspace{1cm} { \LARGE {\bf Spontaneously translational symmetry breaking in the excited states of holographic superconductor}}

\vspace{1.1cm}

Qian Xiang\footnote{xiangq18@.lzu.edu.cn}, Li Zhao\footnote{lizhao@lzu.edu.cn}, Yong-Qiang Wang\footnote{yqwang@lzu.edu.cn, corresponding author}

\vspace{0.7cm}

{\it $^{a}$Lanzhou Center for Theoretical Physics, Key Laboratory of Theoretical Physics of Gansu Province,
	School of Physical Science and Technology, Lanzhou University, Lanzhou 730000, People's Republic of China\\
	$^{b}$Institute of Theoretical Physics $\&$ Research Center of Gravitation, Lanzhou University, Lanzhou 730000, People's Republic of China}

\vspace{0.7cm}
\end{center}
\begin{abstract}
\noindent
We revisit HHH model in [Phys. Rev. Lett. {\bf 101},  031601  (2008)] and extend the ansatz of matter fields to being of depending on a spatial dimension except the holographic direction. Despite homogeneous solutions of ground and excited states, especially for the excited states, there also exists solutions where the translational invariance is broken. It is worth mentioning that no periodic sources are assigned to the matter fields, so the translational symmetry is broken spontaneously. We investigate how the new solutions and the condensates of excited states develop with the change of temperature. Moreover, since this kind of condensate will decrease at certain temperature and eventually vanish at sufficiently low temperature, we also study the relation between this interval and length of lattice. Besides, we compare the free energies of non-translational invariant solutions and those of translational invariance in the HHH model, and find that the free energies of the former situations are lower.  
%---------------------------------------------------------------------------
\end{abstract}
\vspace{5cm}
\pagebreak
\section{Introduction}
As a powerful tool developed from string theroy, the AdS/CFT correspondence, for its tractability, has achieved great success in calculating and modeling strongly coupled theories without gravity in a lower dimensional spacetime. For superconductors, we now have various holographic models of modeling superconductivity which are enlightening in the understanding of mechanism of high temperature superconductivity. The s-wave holographic superconductor was constructed in the seminal papers \cite{Hartnoll:2008vx, Gubser:2008px, Hartnoll:2008kx} where a complex scalar field coupled to a $U(1)$ gauge field in the four dimensional Schwarzschild-AdS black hole was successful in modeling Cooper pairs condensation. Subsequently, the p-wave \cite{Aprile:2010ge, Cai:2013pda, Cai:2013aca, Gubser:2008wv} and d-wave \cite{Chen:2010mk, Benini:2010pr, Kim:2013oba} holographic superconductor were also established by substituting other fields for the complex scalar field. Moreover, other concrete phenomena relating to superconductivity, for instance, Josephson junction \cite{Horowitz:2011dz, Wang:2011rva, Siani:2011uj, Wang:2011ri, Wang:2012yj, Cai:2013sua, Li:2014xia, Liu:2015zca, Hu:2015dnl, Wang:2016jov, Kiczek:2019lmz} was also studied in holographic approach.

However, most of the early studies of holographic superconductor are spatially translational invariant, in which the modeled charged particles would have nowhere to dissipate their momentum resulting a delta function at zero frequency in the real part of optical conductivity even in the normal state, causing infinity in DC conductivity. While in real materials, such symmetry can be broken by lattice. Thus, to break the translational invariance and to recover real experiments is vital in holographic superconductor. In the attempts of modeling such crucial structure, pioneering work was done in \cite{Horowitz:2012ky, Horowitz:2012gs, Horowitz:2013jaa}, where the lattice structure was constructed by adding a periodic source for scalar field on the conformal boundary; the periodicity carried by scalar field was then imprinted into the bulk spacetime, breaking the translational symmetry, namely, scalar lattice. According to the matter that was added to the periodic source, ionic lattice which the periodicity was carried by chemical potential was constructed in \cite{Hartnoll:2012rj}.

However, due to the idea of imprinting periodic structure of matter fields into the spacetime, early attempts fell into technical difficulties of solving partial differential equations (PDEs) embodied in spacetime metric. Afterwards, great simplification was done by homogeneous model, which made the spacetime metric depend on holographic direction only. To simplify numerical works, Q-lattice and helical lattice all involve explicit periodical and non-translational invariant sources while preserving homogeneity of the spacetime metric \cite{Hartnoll:2016apf}. For Q-lattice \cite{Donos:2013eha, Donos:2014uba}, the model gives the modeling phase order, the scalar field, a sourced ansatz which can be viewed as arising from two scalar fields with the same mass and a sinusoidal periodic spatial dependence in a same conformal field boundary direction that is shifted by a phase $\pi/2k$. The helical lattice \cite{Donos:2012js, Donos:2014oha} is produced by a source, dual to a vectorial operator, of a $U(1)$ field which breaks the translational invariance on the conformal field boundary while maintaining invariant under non-abelian Bianchi $\textrm{VII}_0$ symmetry algebra. These models have all reduced the complex numerical study of holography to a simplification that involved ordinary differential equations (ODEs). Except for giving a periodic source, linear axion model \cite{Andrade:2013gsa, Gouteraux:2014hca, Taylor:2014tka, Kim:2014bza} has also realized momentum dissipation, avoiding complex PDEs by exploiting a shift symmetry of the massless scalar field while the source of it is required to be linear in the boundary coordinates. In addition, according to holographic dictionary, to violate the conservation of energy-momentum in the dual field theory can also be realized by giving graviton a mass because it breaks the diffeomorphism invariance in the gravitational theory; as a simplification of avoiding multiple complex PDEs. Since the microscopic details of lattice in the dual field theory are not fully understood therefore, massive gravity model can be viewed as a coarse-grained description of bulk lattice or impurities \cite{Davison:2013jba, Li:2019qkt}. In addition, as dRGT nonlinear massive gravity model is proven to be immune from Boulware-Deser ghost, it has been widely study in holography in \cite{Li:2019qkt, Vegh:2013sk, Zeng:2014uoa}.

Another ingredient in this work are the excited state of holographic superconductor solutions where in \cite{Horowitz:2008bn}, these states are assumed to describe new bound states between quasi-particles. Recent studies in \cite{Wang:2019caf, Wang:2019vaq} have presented a family of solutions in excited state where the profiles of scalar field solutions have multiple nodes along the holographic direction, while they also show additional peaks and poles in optical conductivity. Semi-analytical studies toward the excited state are presented in \cite{Qiao:2020fiv}, while nonequilibrium process of these states have been studied in detail in \cite{Li:2020omw}. Moreover, they have been generalized to the framework of dRGT massive gravity model in \cite{Xiang:2020ugu} as well. 
In this work, we present a family of new solutions of the excited states that do not have translational symmetry. Since there are not periodic sources added into the matter fields thus this kind of symmetry breaking can be viewed as happen spontaneously.
Via the spontaneous mechanism of breaking translational invariance, the results found in this work can provide a physical process of condensate of excited state forming at $T_\textrm{c}$ while vanishing at sufficiently low temperature. Besides, the stabilities of these non-translational invariant meta-stable states are also studied.

Our work is arranged as follows: the holographic setup is given at section \ref{set up}; the numerical solutions of matter fields and details of condensate and temperature are given at subsections \ref{section 3.1}, \ref{section 3.2}; we also analyze the stability of our model at subsection \ref{section 3.3}; brief discussion and conclusion are arranged at section \ref{section 4}.
%---------------------------------------------------------------------------
\section{Holographic setup}\label{set up}
The bulk action is read as follow where a Maxwell field and a charged complex scalar field are coupled in the Einstein gravity with a negative cosmological constant, $\Lambda=-3/\ell^{2}$ where the $\ell$ is the length scale of AdS$_{3+1}$ spacetime.

\be
\mathcal{S}= \frac{1}{16\pi G}\int \mathrm{d}^4x \left[R+\frac{6}{\ell^{2}}-\frac{1}{4}F^{\mu\nu}F_{\mu\nu}-(\mathcal{D}_\mu \psi)(\mathcal{D}^\mu \psi)^*-m^{2}\psi \psi^*\right]. \label{Lagdensity}
\ee

The action we use is just the minimal gravitational action introduced in \cite{Hartnoll:2008vx} with the gauge covariant derivative $\mathcal{D}_\mu=\nabla_\mu-\textrm{i}q A_\mu\psi$, where a abelian gauge field $A_\mu$ minimally couples a scalar field $\psi$ with mass $m$ and charge $q$. The field strength of the $U(1)$ gauge field is represented by $F_{\mu\nu}=\partial_{\mu}A_{\nu}-\partial_{\nu}A_{\mu}$.

Since we will work in the probe limit approximation, we adopt $q\rightarrow \infty$ with the following scaling transformations.

\be
A \rightarrow A/q, ~~~~\psi \rightarrow \psi/q.
\ee

Under the probe limit setting, because of a $1/q^2$ in front of the matter fields of the lagrangian density (\ref{Lagdensity}), the gravity is decoupled from the matter fields and thus the solution of Einstein equation is just the Schwarzschild anti-de Sitter black hole:

\begin{eqnarray} \label{metric}
  \textrm{d}s^{2} = -f(r)\textrm{d}t^{2}+\frac{\textrm{d}r^{2}}{f(r)}+r^{2}(\textrm{d}x^{2}+\textrm{d}y^{2}),
\end{eqnarray}

Where $f(r)=\frac{r^2}{\ell^2}(1-r_\textrm{h}^3/r^3)$, and $r_\textrm{h}$ is the radius of the event horizon, which determines the Hawking temperature of the black hole:

\be\label{T}
T=\frac{1}{4\pi}\frac{\textrm{d}f}{\textrm{d}r}\bigg|_{r=r_\textrm{h}}=\frac{3r_\textrm{h}}{4\pi \ell^{2}}.
\ee
For simplicity, hereafter, we will numerically set $\ell=1$. And the non-backreacting gauge field and the scalar field determined by the coupled field equations read

\begin{eqnarray}
\centering
(\nabla_{\mu}-\textrm{i}A_{\mu})(\nabla^{\mu}-\textrm{i}A^{\mu})\psi-m^{2}\psi&=&0,\label{scalarequ}\\
\nabla_{\mu}F^{\mu\nu}-\textrm{i}[\psi^{\ast}(\nabla^{\nu}-\textrm{i}A^{\nu})\psi-\psi(\nabla^{\nu}+\textrm{i}A^{\nu})\psi^{\ast}]&=&0.\label{maxwellequ}
\end{eqnarray}

So far, the above settings are just the non-backreacting s-wave holographic superconductor if the spacetime and matter fields are only determined by radius coordinate $r$. Since we want to find solutions that are non-trivially dependent on the conformal boundary coordinate, say $x$ direction, and thus breaks the translational symmetry. We adopt the following ansatz

\be \label{ansatz}
	\psi = \psi(r,x), ~~~~A=\phi(r,x)\textrm{d}t.
\ee

Therefore, by considering the specific direction $x$ on the boundary these two matter fields satisfy
\begin{eqnarray}
	\frac{1}{r^2f}\psi_{xx}+\psi'' + \left(\frac{f'}{f} + \frac{2}{r}\right) \psi' +\frac{\phi^2}{f^2}\psi - \frac{m^2}{ f} \psi &=& 0 \,,\label{eom1}\\
	\frac{1}{r^2f}\phi_{xx}+\phi'' + \frac{2}{r} \phi' - \frac{2\psi^2}{f} \phi &=& 0.\label{eom2}
\end{eqnarray}
Where, the prime denotes derivative on radius $r$, while subscript $x$ means derivative on $x$ direction. Compare (\ref{eom1}) and (\ref{eom2}) with equations of motions derived, for example, from \cite{Hartnoll:2008vx} one can immediately notice there are two second rank derivative terms of both $\psi$ and $\phi$ according to $x$. Moreover, the matter field solutions in s-wave holographic superconductor without backreaction also satisfy our equations of motions since the first and second derivatives on $x$ of the homogeneous matter fields solutions are zero. At $r\rightarrow \infty$ boundary, the asymptotic behaviours of the matter fields are

\begin{eqnarray} \label{scalar}
\psi(r,x)& \stackrel{r\rightarrow \infty}{\longrightarrow}&\frac{\psi^{(1)}(x)}{r^{\Delta_-}}+\frac{\psi^{(2)}(x)}{r^{\Delta_+}}+ \cdots,\\
\phi(r,x)& \stackrel{r\rightarrow \infty}{\longrightarrow}&\mu - \frac{\rho(x)}{r} + \cdots,\label{maxwell}
\end{eqnarray}
where, according to AdS/CFT dictionary, $\Delta_{\pm}=(3\pm\sqrt{9+ 4 m^2})/2$ are interpreted as the scaling dimensions of the dual field theory operators. As long as one of the operators, say $\langle \ocal_{\Delta_+}\rangle$, acquires a non-vanishing vacuum expectation value, the $\psi^{(2)}(x)$ is dual to the condensate value with $\psi^{(1)}(x)$ being its source. Since we require the translational symmetry on the boundary to be broken spontaneously we choose $\psi^{(1)}(x)=0$. And for simplicity, we also choose $m^2=-2$. The chemical potential and the charge density are represented by $\mu$ and $\rho(x)$, respectively. In scalar lattice model, the periodicity of charge density is introduced by adding a periodic source for the neutral scalar operator, while keeping the chemical potential constant. Here, we also keep chemical potential constant, but without a source. Also, we will conduct a coordinate transformation for the sake of simplicity of numerical calculation, which reads
\be
z=r_\textrm{h}/r,~~~~z_\infty=0,~~~~z_\textrm{h}=1.
\ee
Such that the integration region will be $0\leq z \leq 1$, while in the $z$ direction the boundary conditions we choose are as follow
\be\label{BC1}
\psi(0,x)=0,~~~~\phi(0,x)=\mu,~~~~\phi(z_\textrm{h},x)=0.
\ee

It is easy to notice that this is just the boundary condition imposed on s-wave holographic superconductor with translational symmetry, where below critical temperature the scalar hair can live outside the black hole while vanishing on the conformal boundary; the Abelian field entirely absorbed near the horizon, while asymptotically approaching $\mu$ on the boundary. When the temperature $T$ is higher than critical temperature $T_\textrm{c}$, the matter fields vanish, thus illustrating a scenario without superconductivity.

In the following, by means of an iterative process based on Newton-Raphson method, we will show that within this simple set up there does exist solutions of excited states with broken translational symmetry.

%On the $x$ direction of the conformal boundary, \textcolor{red}{\st{since we also want periodicity embodied in the matter fields}}, both Dirac boundary conditions and Newman boundary conditions will be, respectively, imposed on $x=0,x=L$ and $x=L/2$, which read

%\begin{eqnarray}\label{BC2}
%\psi(z,0)=\psi(z,L),~~~~\partial \psi(z,x)|_{x=L/2}=0;\\
%\phi(z,0)=\phi(z,L),~~~~\partial \phi(z,x)|_{x=L/2}=0. \label{BC3}
%\end{eqnarray}
%Where $L$ is a constant that can be interpreted as length scale of a ``lattice".

\section{Numerical results}
\begin{figure}[h]
\centering
\includegraphics[height=.3\textheight,width=.3\textheight, angle =0]{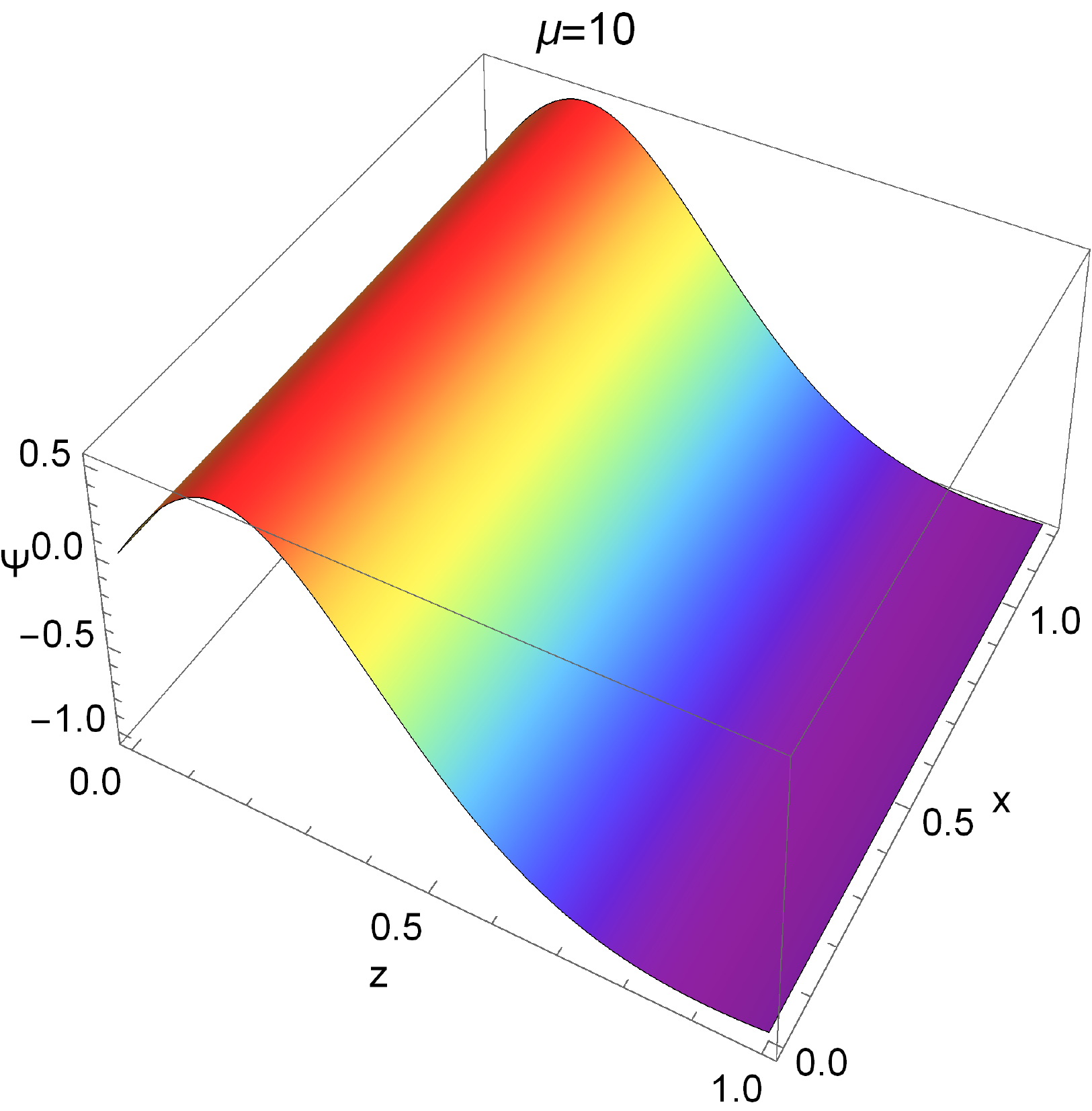}
	\caption{Example of scalar field in first excited state with translational symmetry below $T_\textrm{c}$. }\label{none}
\end{figure}

Before giving the results of excited states with broken translational symmetry, we would like to introduce how the excited states are solved out at first. Using an iterative process by means of Newton-Raphson method, the scalar field solution of $n$-th excited state that possesses $n$ nodes along the radius coordinate will be given a good initial guess that also has $n$ nodes. As an example, the profile of homogeneous scalar field solution of the first excited state is shown in Fig. \ref{none} by giving an initial guess that has one node along $z$ coordinate.

Based on this idea, we will give an initial guess for solution with broken symmetry on $x$ coordinate as well. For example, below $T_\textrm{c}$, the initial guess on the conformal boundary for the scalar field can be set as $\psi_{\textrm{initial}}(0,x)=\beta \textrm{cos}(2\pi x/L)$, where $\beta$ is a constant. Here, we define $0 \leq x\leq L$ as a lattice's length, where $L$ can be interpreted as a length scale of the lattice.
%%%%---------------------------------------------------------------------------------------------------
%%%%---------------------------------------------------------------------------------------------------%%%%---------------------------------------------------------------------------------------------------
\subsection{Solutions with broken translational symmetry}\label{section 3.1}
\begin{figure}[t]
\includegraphics[height=.3\textheight,width=.3\textheight, angle =0]{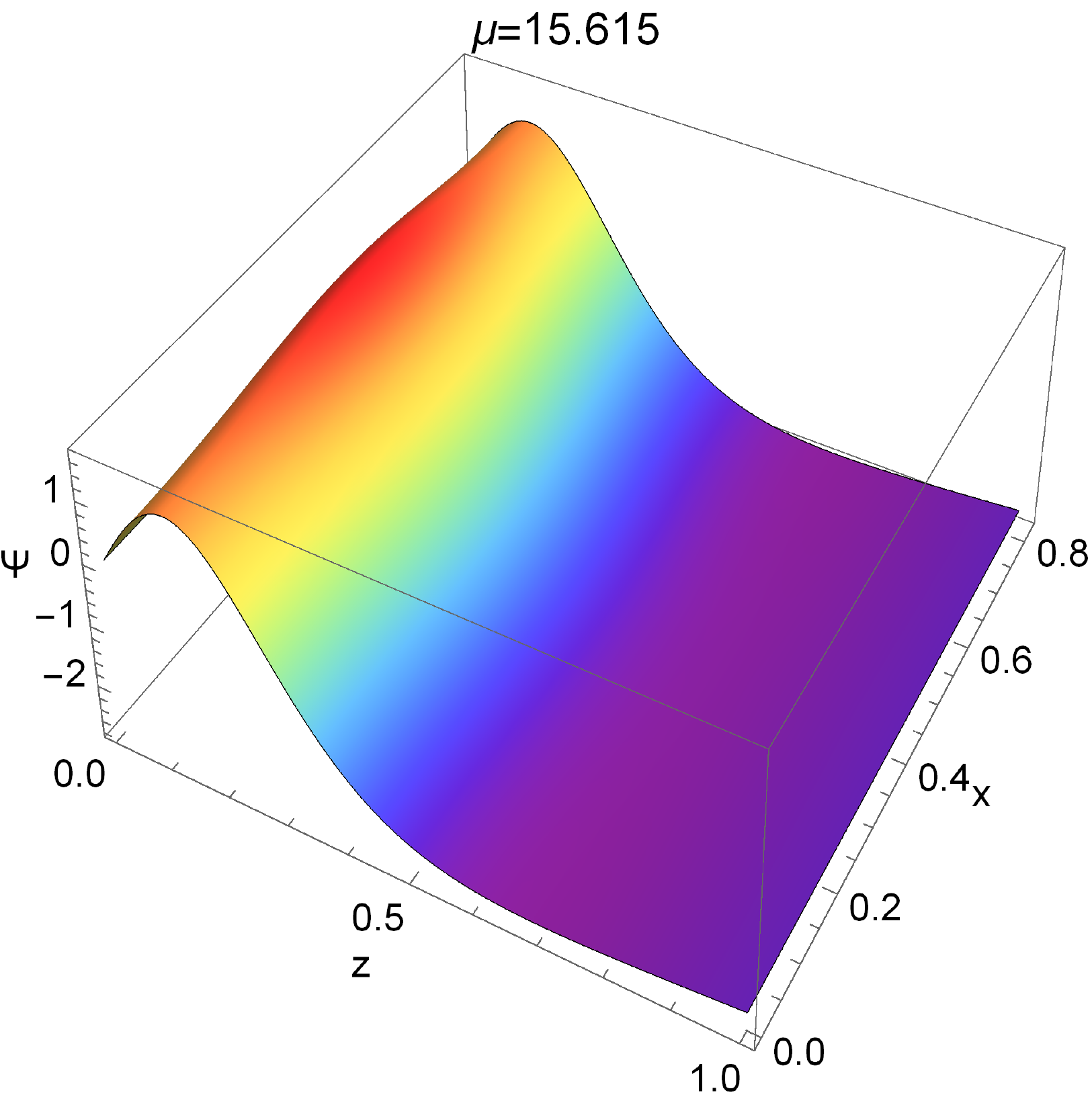}
\includegraphics[height=.3\textheight,width=.3\textheight, angle =0]{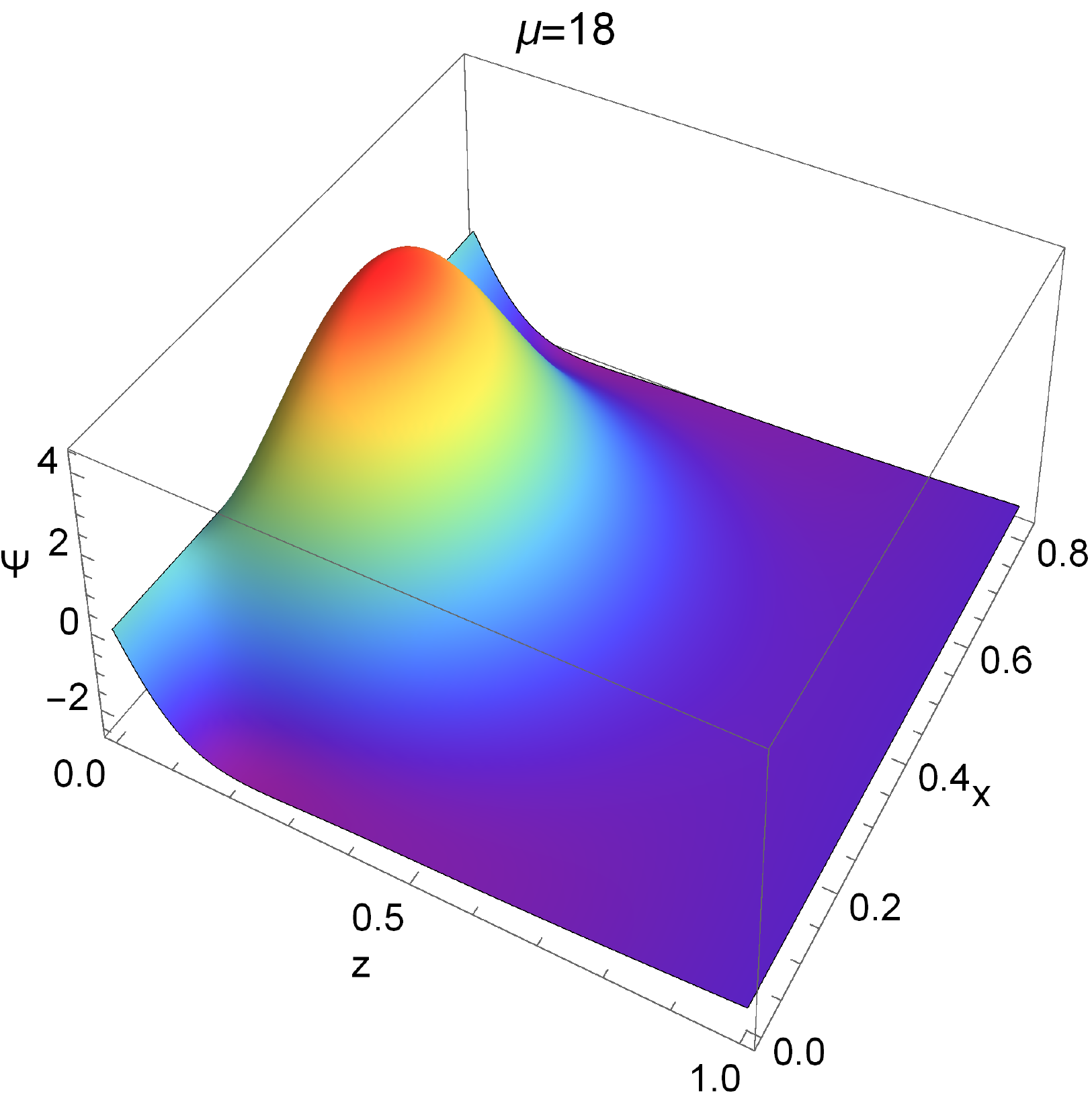}
\centering
\includegraphics[height=.3\textheight,width=.3\textheight, angle =0]{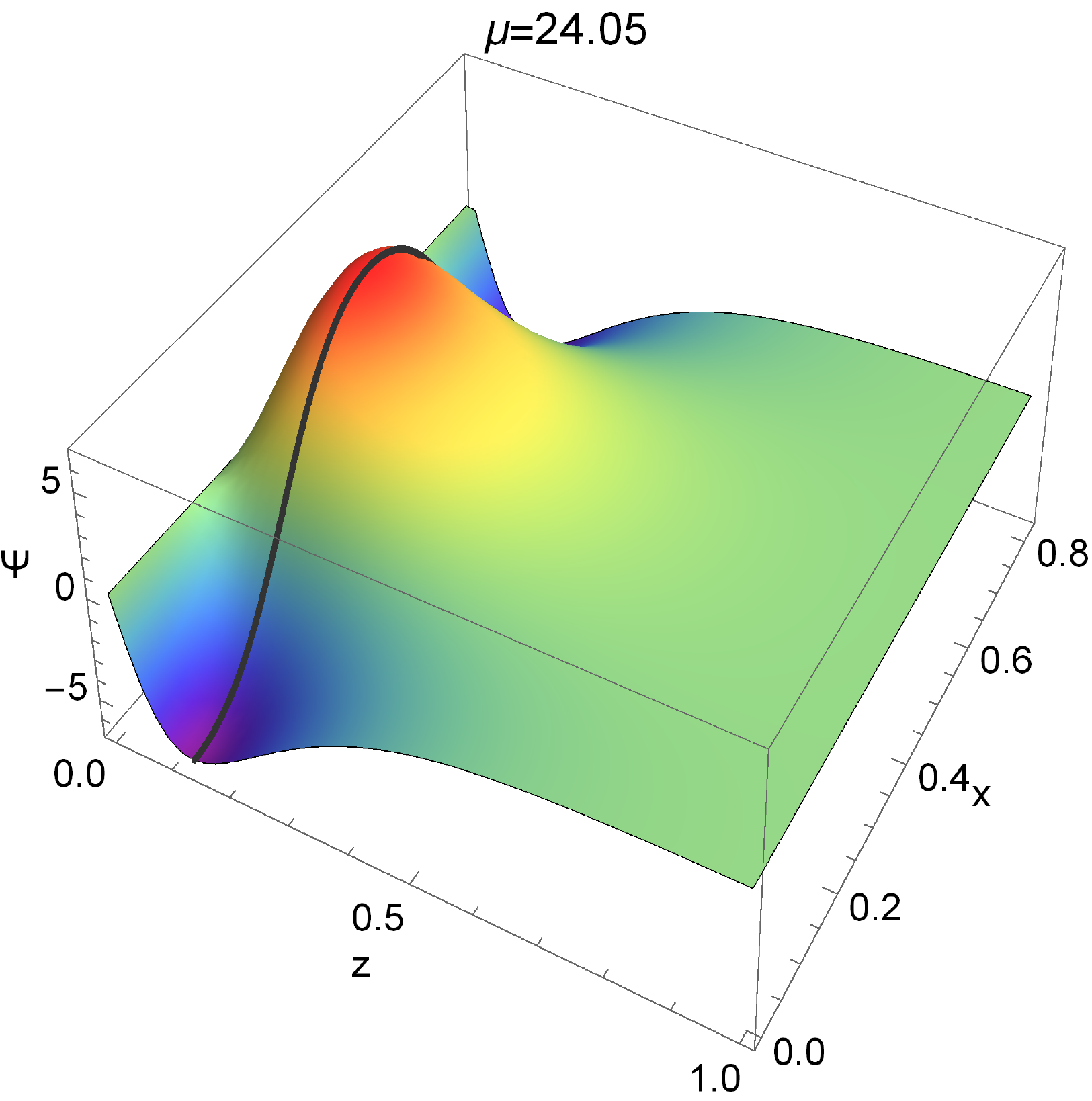}
	\caption{Example of scalar field solution in the first excited state with broken translational symmetry initial value developing from $\mu=15.615$ to $\mu=24.05$ with length $L=0.825$.}\label{first broken psi}
\end{figure}

In Fig. \ref{first broken psi} we present example of the scalar field solutions in the first excited states where we fix $L=0.825$. Subfigures on upper left, upper right and under middle correspond to $\mu=15.615,~\mu=18$ and $\mu=24.05$, respectively. Recall that in a probe limit approximation, temperature is proportional to $\mu^{-1}$, thus these subfigures show how the solution develops from high temperature to low temperature. 

From the figure, we can see that below $T_\textrm{c}$ the tendency of the broken translational symmetry is tiny when $\mu$ is small and mainly concentrates on the ridge between $z=0$ and the first node. After further cooling the system, the tendency of the broken region becomes larger, developing to a peak at $x=L/2$, while the values near $x=0$ and $x=L$ decrease. When the temperature is sufficiently low the peak continues to grow and the values at $x=0,~x=L$ develop to two valleys, forming almost a cosine function at $z\approx 0.1$ (black line in Fig.\ref{first broken psi}) right between $z=0$ and the original first node, while the scalar field vanishes around the horizon indicating that there is no condensate.

\begin{figure}[h]
\includegraphics[height=.3\textheight,width=.3\textheight, angle =0]{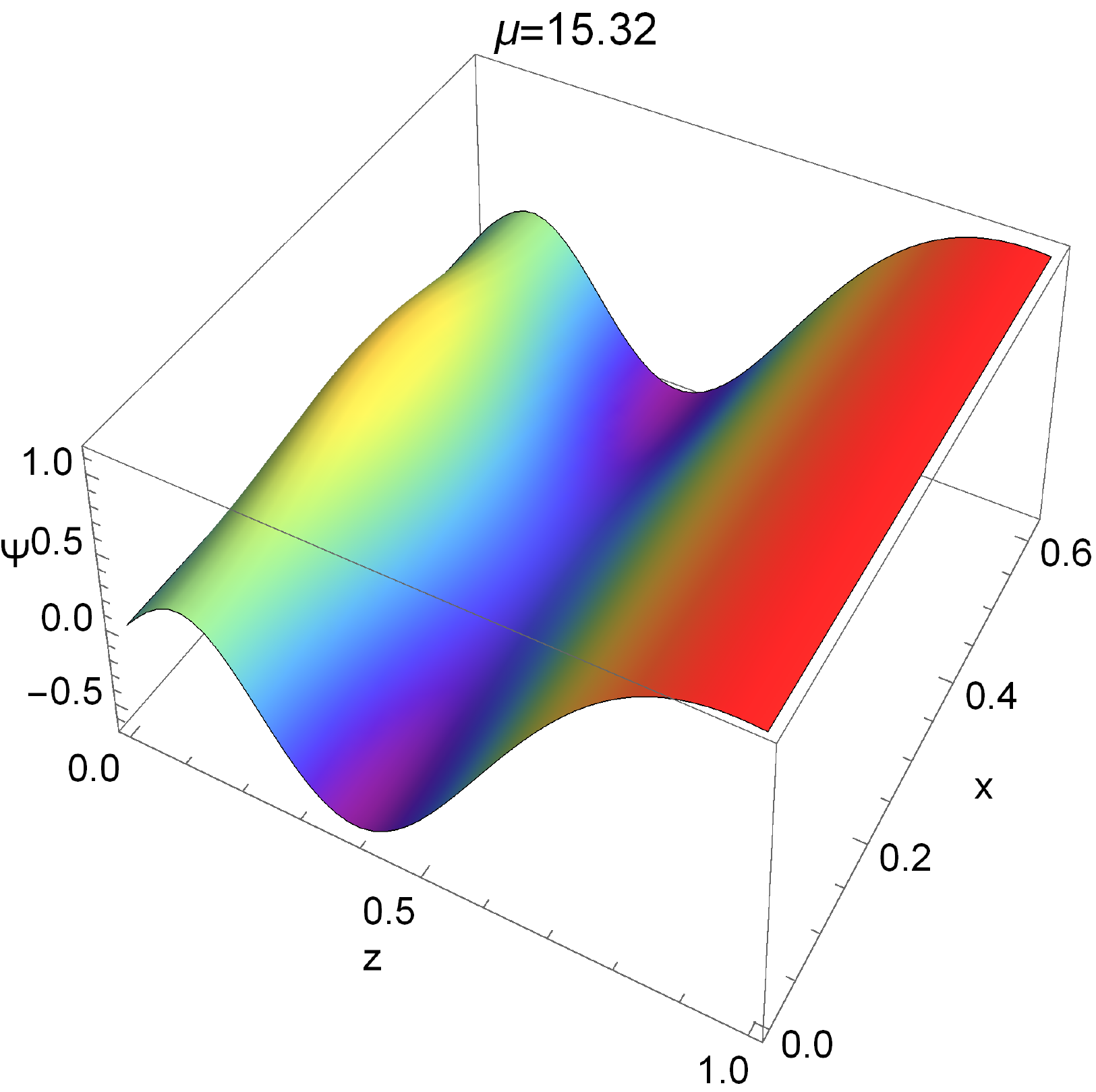}
\includegraphics[height=.3\textheight,width=.3\textheight, angle =0]{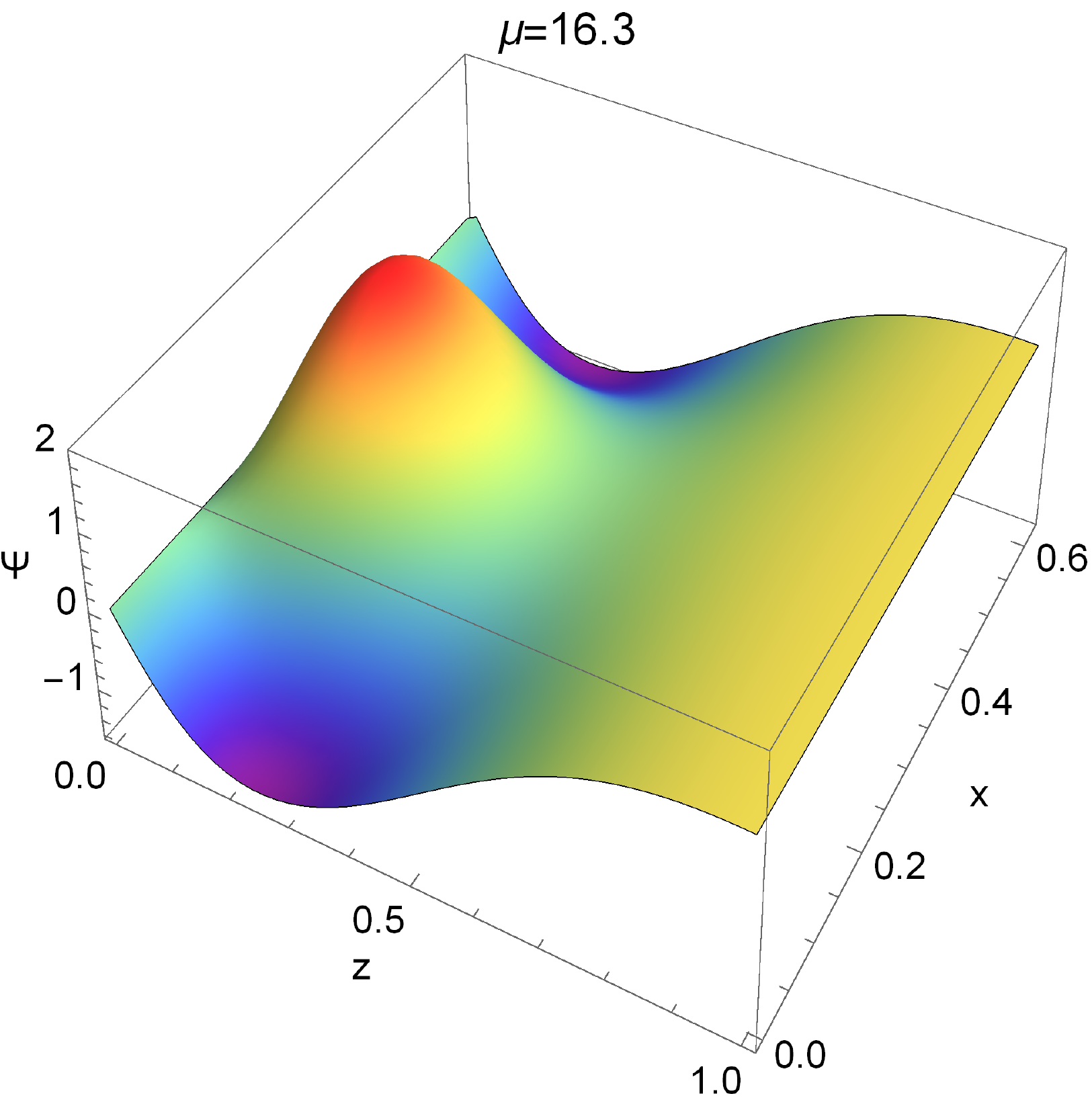}
\centering
\includegraphics[height=.3\textheight,width=.3\textheight, angle =0]{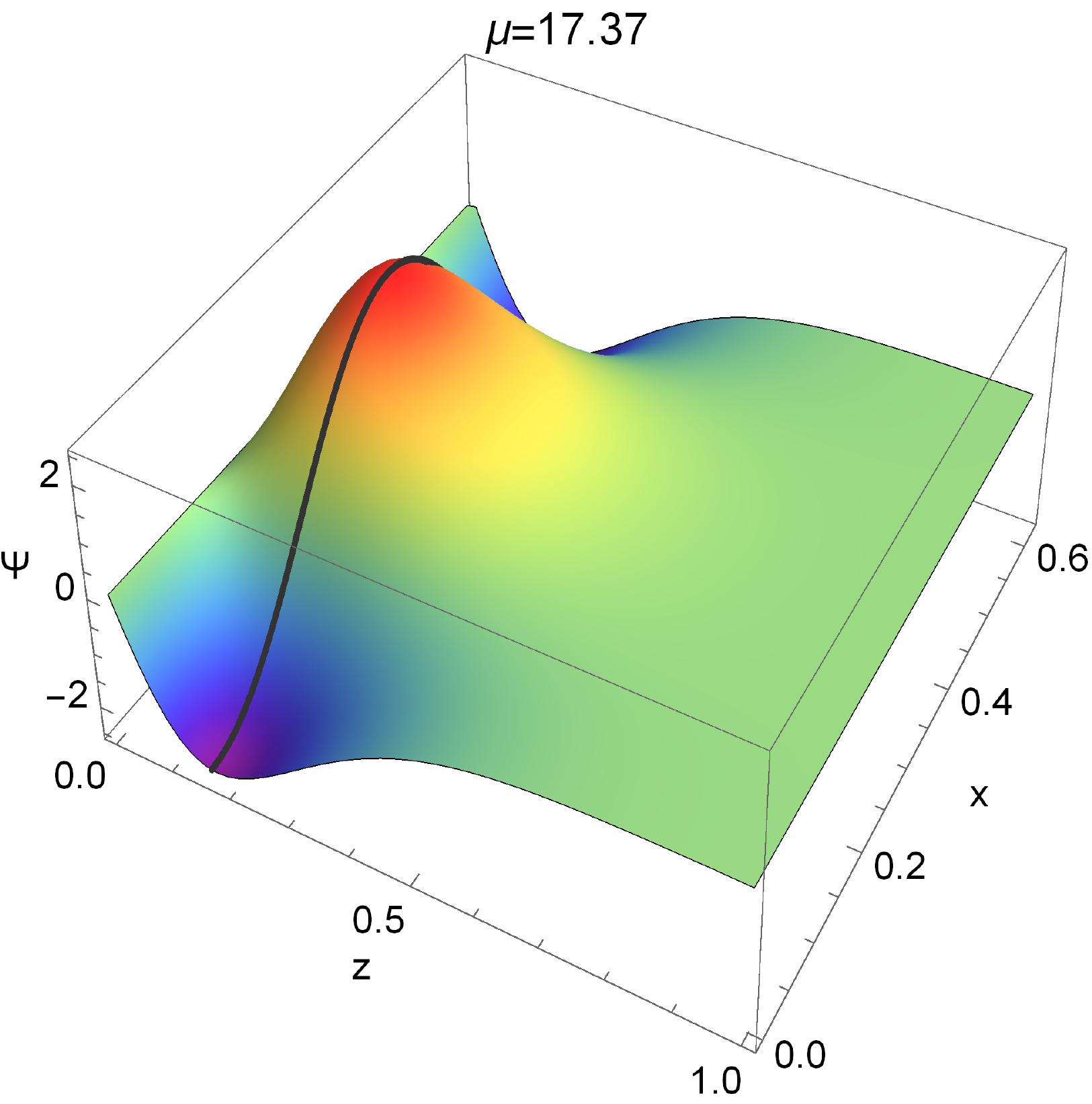}
	\caption{Example of scalar field solution in the second excited state with broken translational symmetry initial value developing from $\mu=15.32$ to $\mu=17.37$ with length $L=0.625$.}\label{second broken psi}
\end{figure}

In Fig. \ref{second broken psi}, we show the process of scalar field solution with broken translational symmetry initial value in the second excited state developing from relatively high temperature to sufficiently low temperature, where the lattice scale is fix at $L=0.625$, and subfigures on upper left, upper right and under middle correspond to $\mu=15.32,~\mu=16.3$ and $\mu=17.37$, respectively. Similar to the first excite state, the translational symmetry is, firstly, broken spontaneously on the ridge between $z=0$ and the first node. A peak is formed at $x=L/2$ with values on the two edges of $x$ coordinate develop to two valleys. Compare with the first excited state, even though the scalar field solution has two nodes initially, it is interesting that the broken area exists only on the first ridge; when the temperature is sufficiently low, the scalar field profiles of these two excited states develop to an identical pattern where the scalar field vanishes near the horizon and thus no condensate. For the convenience of discussion, we define the chemical potential where translational symmetry of the scalar field initially breaks as $\mu_\textrm{d}$, while the chemical potential where condensate vanishes as $\mu_\textrm{v}$; the corresponding temperatures are thus marked as $T_\textrm{d}$ and $T_\textrm{v}$, respectively.

\subsection{Condensate and chemical potential}\label{section 3.2}
\begin{figure}[t]
\includegraphics[height=.24\textheight,width=.33\textheight, angle =0]{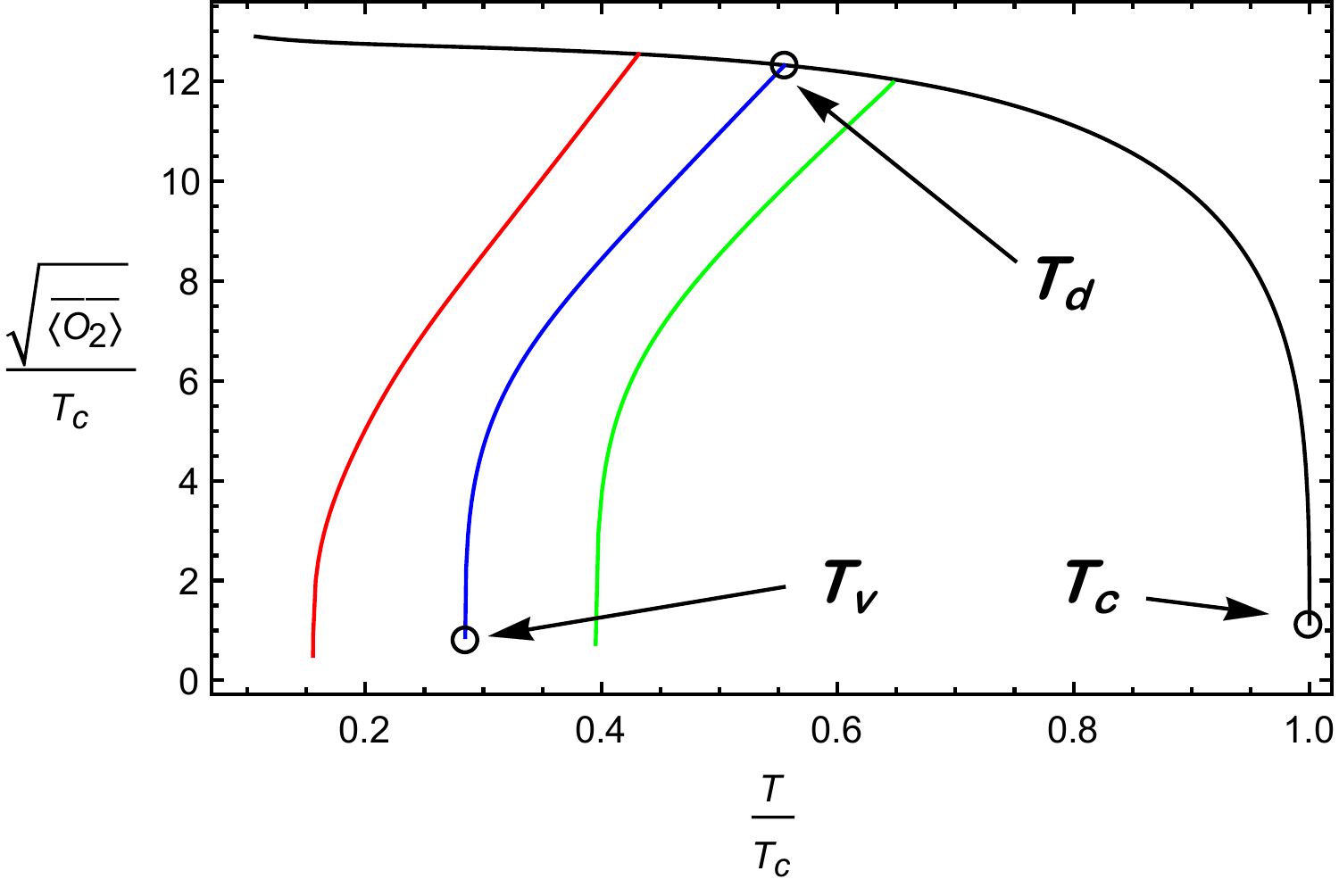}
\includegraphics[height=.24\textheight,width=.33\textheight, angle =0]{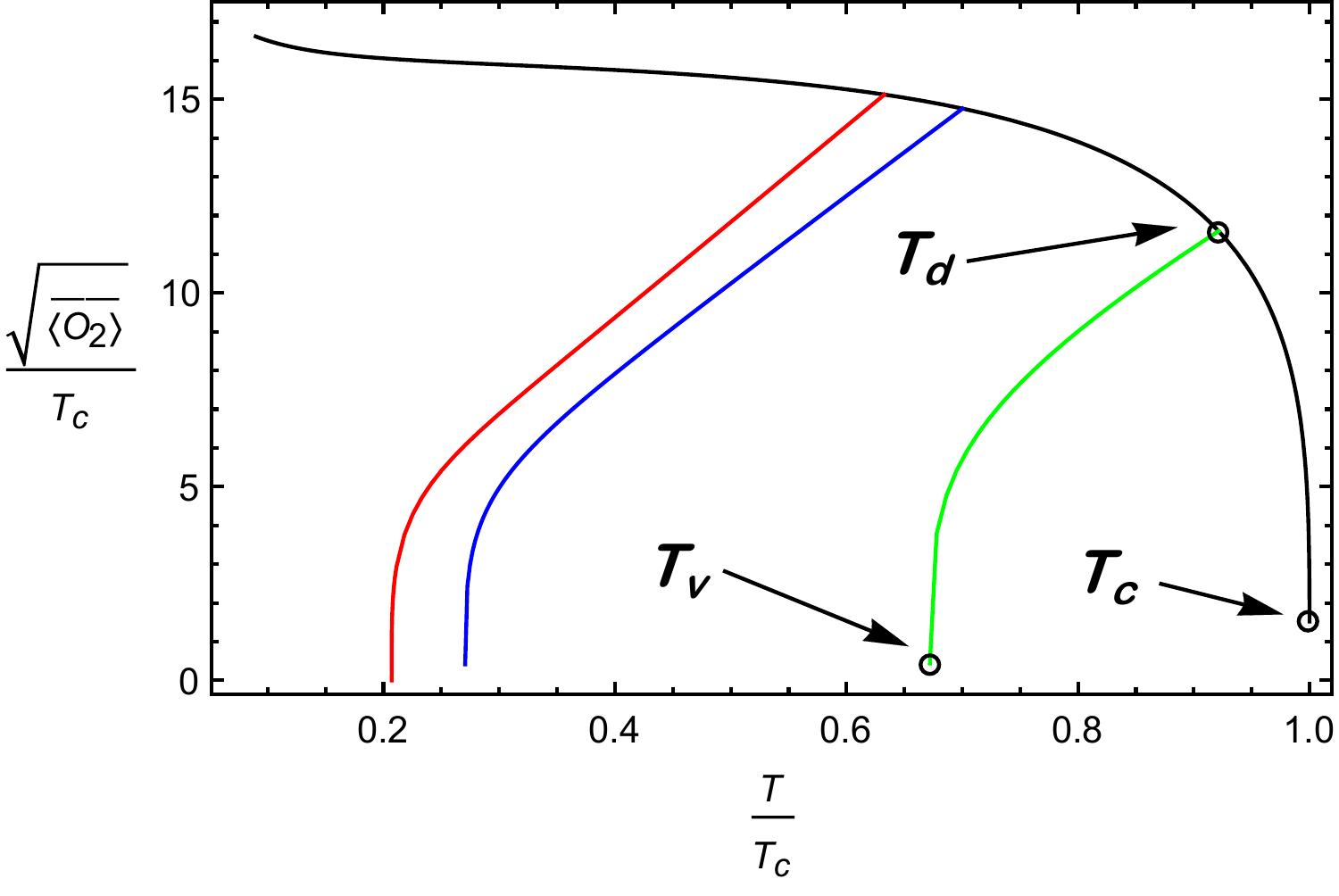}
	\caption{Consensate in the first (left) and second (right) excited states, where black lines correspond to situations with translational symmetry. Red, blue and green lines in the left panel correspond to $L=0.65,~L=0.825$ and $L=0.95$, respectively. Red, blue and green lines in the right panel correspond to $L=0.455,~L=0.5$ and $L=0.625$, respectively.}\label{broken con}
\end{figure}

In Fig. \ref{broken con}, we show condensate in first (left panel) and second (right panel) excited states, where black lines correspond to these states with translational symmetry while red, blue and green lines are results under different $L$ and without the symmetry. Since we now have a hairy black hole with rippled matter fields, the condensates are read off from the mean expectation value of the scalar field, namely,
\be
\overline{\langle \ocal_2 \rangle}=\overline{\psi^{(2)}(x)}\sqrt{2}.
\ee
Where the results shown in black lines are identical to those of excited states presented in previous work \cite{Wang:2019caf} with perfect translational symmetry. The junctions marked with $T_\textrm{d}$ and $T_\textrm{v}$ correspond to the situations in the upper left and under middle subfigures shown in Fig. \ref{first broken psi} and Fig. \ref{second broken psi}.

From the figure, we can clearly see the physics of our model --- below $T_\textrm{c}$, the condensate  forms; after that its value starts to decrease under a temperature $T_\textrm{d}$ due to spontaneously translational symmetry breaking; eventually, when reaching another temperature $T_\textrm{v}$, the condensate vanishes; between $T_\textrm{c}$ and $T_\textrm{v}$ the state is superconducting. In the figure, we also give condensates in the two states with different $L$, where one can find that the corresponding $T_\textrm{d}$ and $T_\textrm{v}$ become higher by increasing $L$. %The process of the mean value of condensate decreases at $T_\textrm{d}$ while vanishes at $T_\textrm{v}$ can also be seen by looking at the amplitute of $\langle \ocal_2 \rangle(x)$. In Fig. \ref{evolution}, we give $\langle \ocal_2 \rangle(x)$ of first excited state with certain $L$ developing from $\mu_\textrm{d}$ to $\mu_\textrm{v}$. $\langle \ocal_2 \rangle(x)$ of different $\mu$, from small to large, are marked by black, orange, purple and red lines, respectively. From the figure, we can see that the mean value of $\langle \ocal_2 \rangle(x)$ is the largest at $\mu=15.62$; when
This phenomenon of condensate forming below $T_\textrm{c}$ while decreasing at a temperature $T_\textrm{d}$ and eventually vanishing at sufficiently low temperature $T_\textrm{v}$ has also been seen in multi-order holographic model. For example, in \cite{Li:2014wca}, condensates are modeled by two scalar fields, and thus each condensate has different critical temperature. When one of the scalar fields, say $\psi_1$, starts to condense, the other is, at that moment, zero. After continuing to lower the temperature, condensate corresponding to $\psi_1$ decreases, while the other starts to form; under a sufficiently low temperature, the condensate corresponding to $\psi_1$ vanishes. In our model, the process of condensate that forms under $T_\textrm{c}$ and eventually vanishes at $T_\textrm{v}$ closely resembles the process of one of the scalar fields in two fields competing model. 
\begin{figure}[t]
\includegraphics[height=.24\textheight,width=.33\textheight, angle =0]{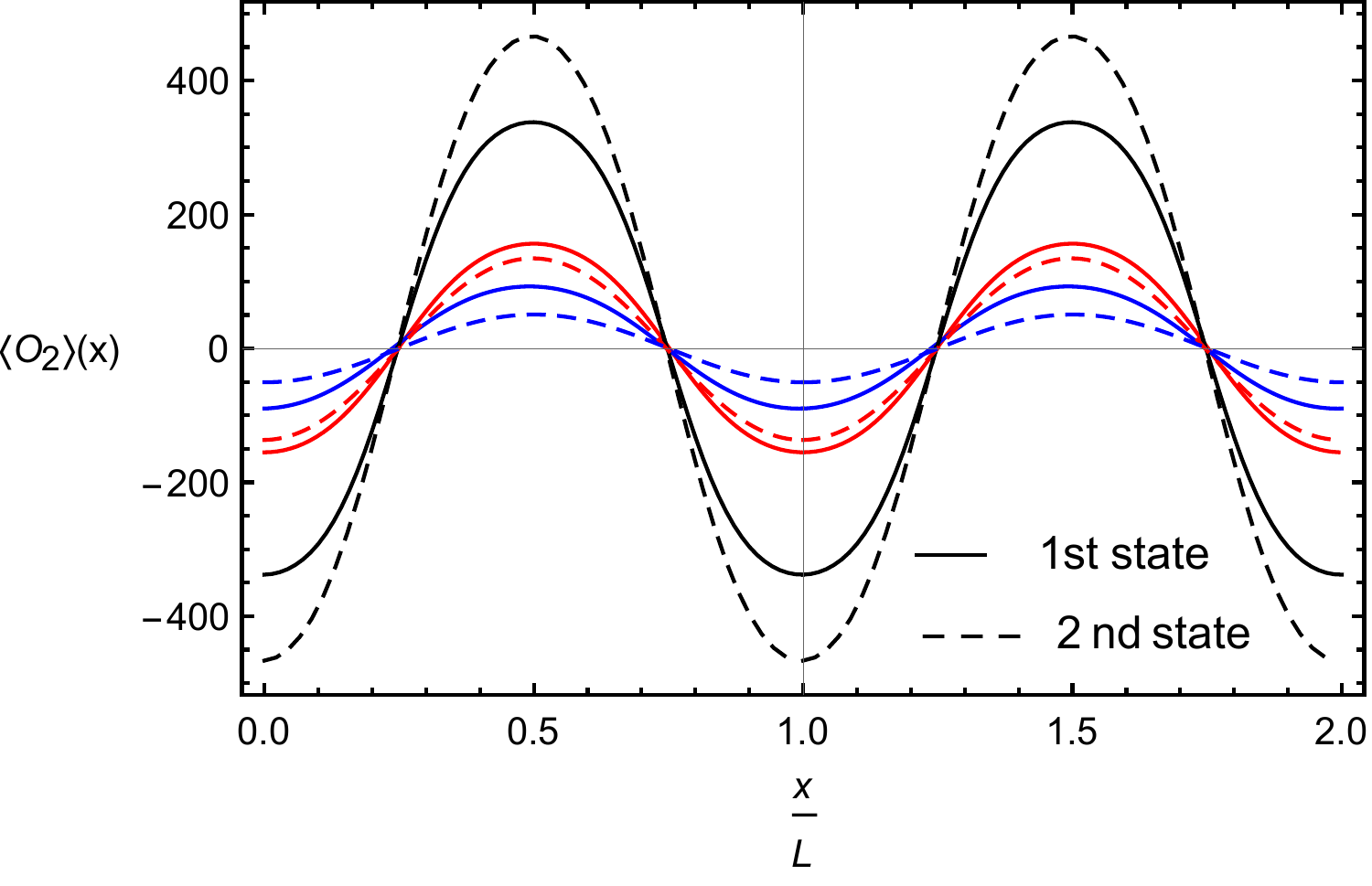}
\includegraphics[height=.24\textheight,width=.33\textheight, angle =0]{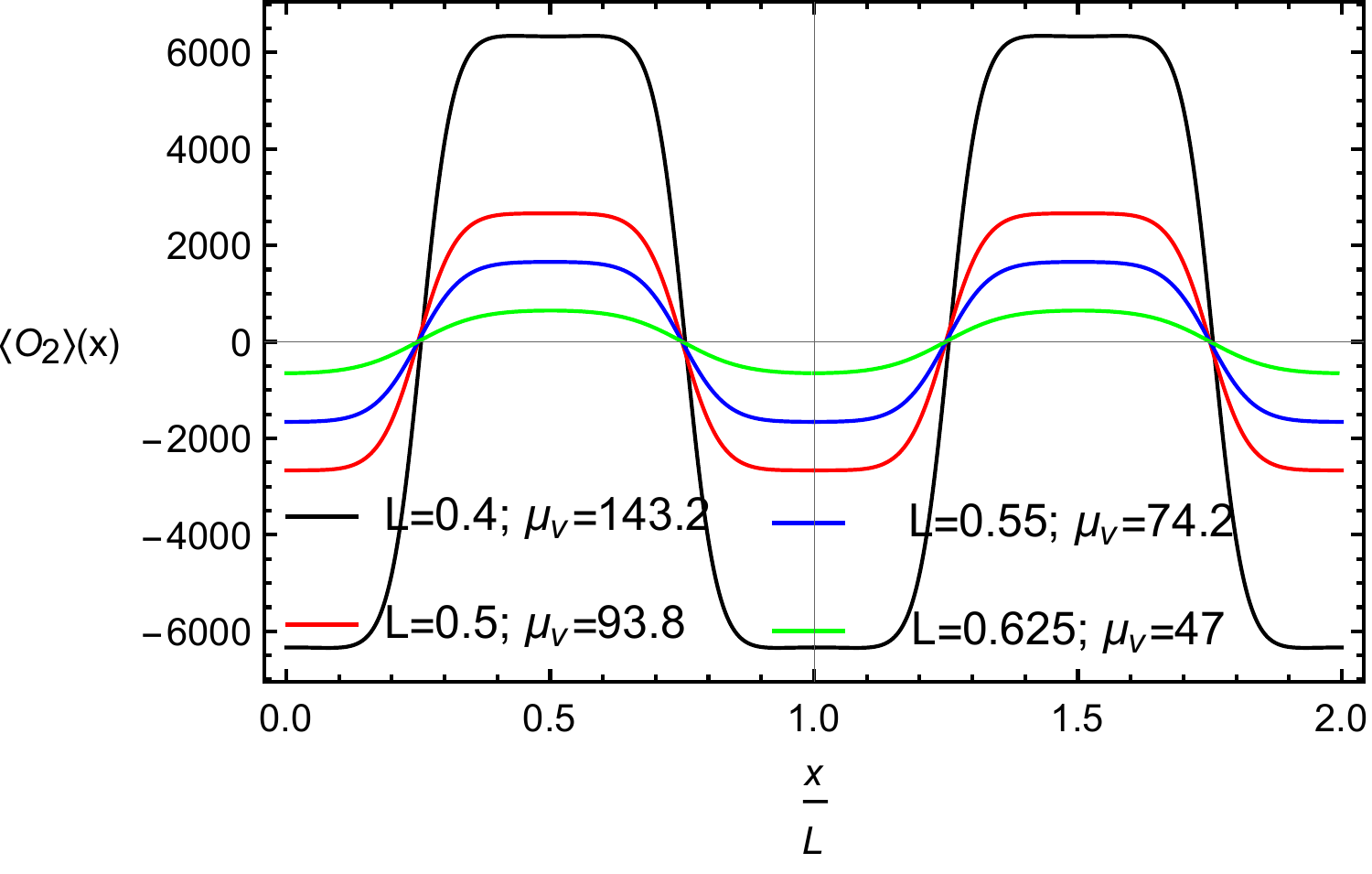}
	\caption{Kink-like condensates $\langle \ocal_2 \rangle (x)$. Solid black, red and blue lines in the left panel correspond to the first excited state with $L=0.7,~\mu_\textrm{v}=34.30$, $L=0.825,~\mu_\textrm{v}=24.05$ and $L=0.925,~\mu_\textrm{v}=19.05$, respectively. Dashed black, red and blue lines in the left panel correspond to the second excited state with $L=0.455,~\mu_\textrm{v}=41.67$, $L=0.555,~\mu_\textrm{v}=25.16$ and $L=0.615,~\mu_\textrm{v}=18.40$, respectively. Solid black, red, blue and green lines in the right panel are $\langle \ocal_2 \rangle (x)$ of the first excited state corresponding to $L=0.4,~\mu_\textrm{v}=143.2$, $L=0.5,~\mu_\textrm{v}=93.8$, $L=0.55,~\mu_\textrm{v}=74.2$ and $L=0.625,~\mu_\textrm{v}=47$, respectively.}\label{kink1}
\end{figure}

\begin{figure}[h]
\centering
\includegraphics[height=.35\textheight,width=.6\textheight, angle =0]{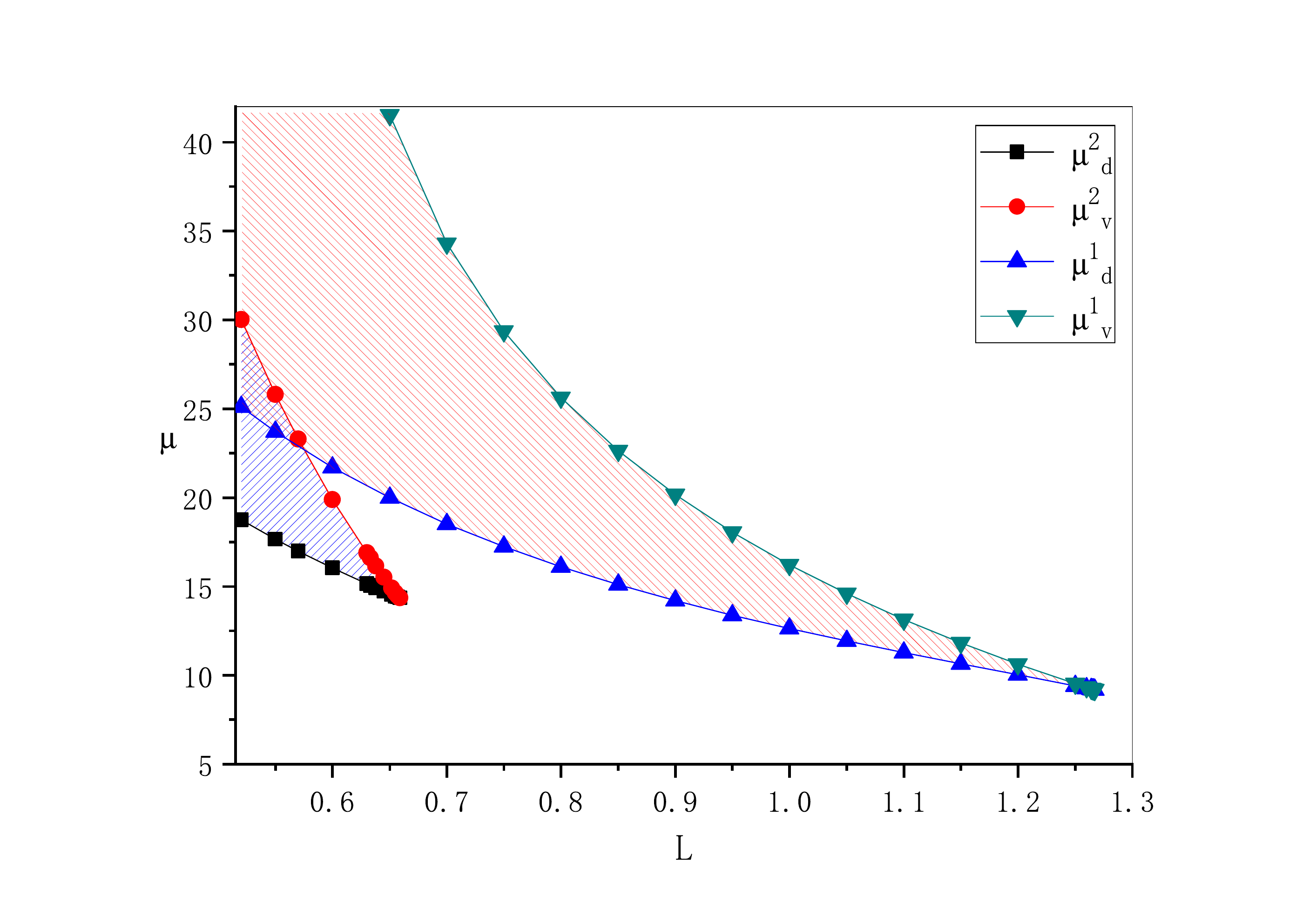}
	\caption{Relations between $\mu_\textrm{d},~\mu_\textrm{v}$ and $L$ in first and second excited states, where black and red lines correspond to $\mu_\textrm{d}$ and $\mu_\textrm{v}$ in the second excited state, while blue and green lines relate to $\mu_\textrm{d}$ and $\mu_\textrm{v}$ in the first excited state.}\label{mu gap}
\end{figure}

Another interesting feature at sufficiently low temperature $T_\textrm{v}$ (or $\mu_\textrm{v}$) is the kink-like solution, as shown in Fig. \ref{kink1}. In the figure, lines of different colours are condensation amplitudes of different length scales. Recall that the rippled condensates are solved out from initial guesses of cosine functions with $L$ being their periods, thus, the figure we show are $\langle \ocal_2 \rangle (x)$ in two adjacent lattices. Where, in the left panel, for both the first (solid lines) and second (dashed lines) excited states, one can find that the amplitudes of these non-superconducting solutions decrease with the increase of $L$. Analogous kink-like solutions can also be found at \cite{Matsumoto:2019ipg, Lan:2017qxm, Xu:2019msl, Correa:2009xa}, where, as the previous holographic studies illustrated, these inhomogeneous solutions are important results from microscopic models like the BCS theory and the Gross-Neveu model, and conventional phenomenological Ginzburg-Landau theory with higher-derivative interactions terms, to which holographic models must recover. Except for the left panel where $L$ are relatively large (we will see that $L$ have limits in Fig. \ref{mu gap}), $\langle \ocal_2 \rangle (x)$ of the first excited state shown in the right panel that have smaller $L$ qualitatively resemble kink crystalline condensate refered in \cite{Matsumoto:2019ipg}. Where, in the right panel, with the decrease of $L$, the peaks of $cos$-like functions grow higher and the smooth peaks become flat.

%Except for the above solutions of two nodes living in one lattice space, our model can also qualitatively recover solutions of more nodes living in one lattice space as well. The example solved out from the first excited state is given in the right panel of Fig. \ref{kink1}. Where, in the figure, with the decrease of $L$, the peaks of $sin$-like functions become higher and the smooth peak becomes flat. This result is qualitatively similar to kink crystalline condensate refered in \cite{Matsumoto:2019ipg}. 

\begin{figure}[t]
\centering
\includegraphics[height=.24\textheight,width=.33\textheight, angle =0]{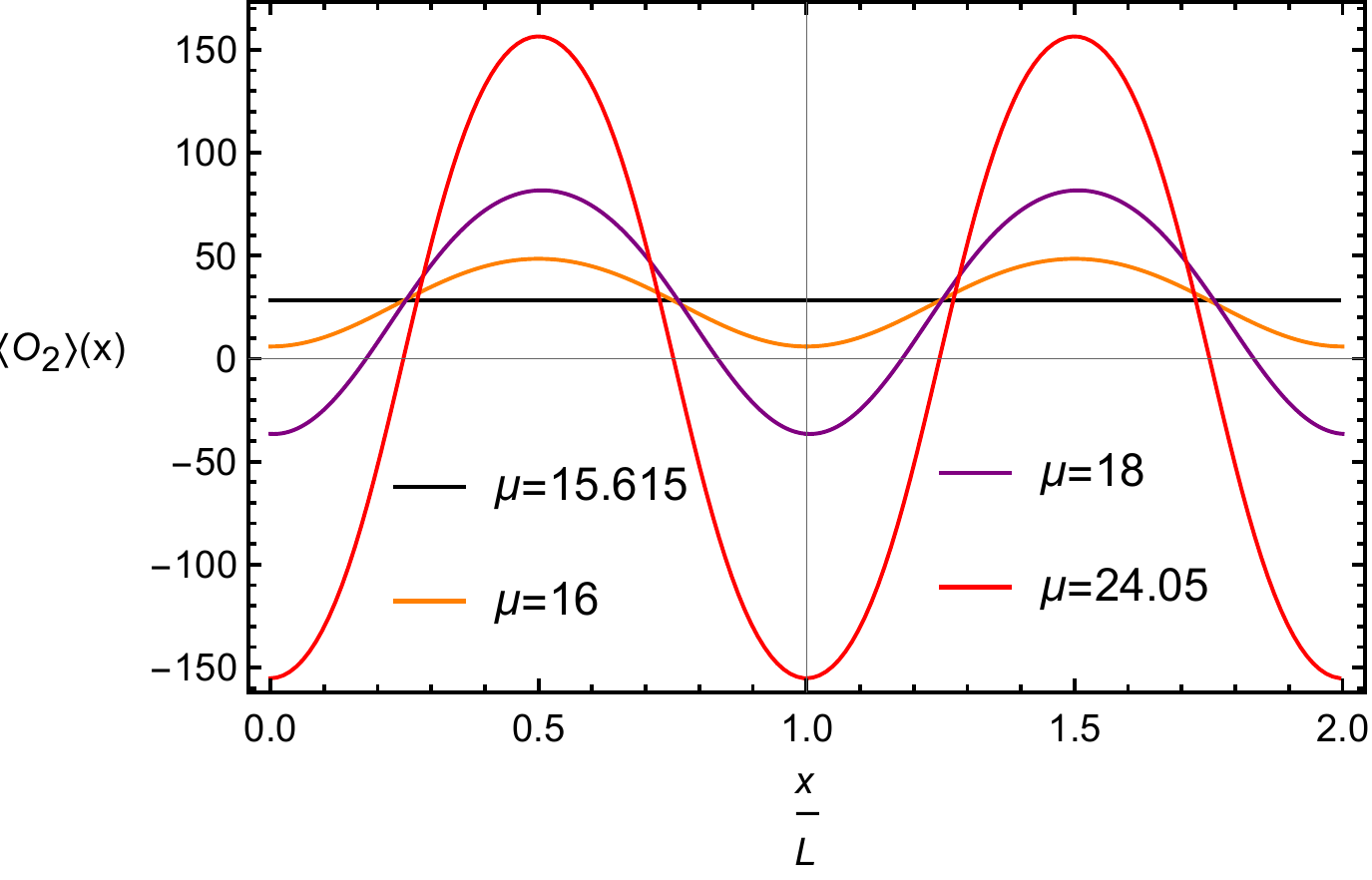}
	\caption{Process of $\langle \ocal_2 \rangle(x)$ of first excited state developing from relatively high temperature to sufficiently low temperature with $L=0.825$, where lines of different colours correspond to different temperatures. $\mu=15.62~(\text{black}),~16 ~(\text{orange}),~18~(\text{purple}),~24.05~(\text{red})$.}\label{evolution}
\end{figure}

A common feature of the previous studies is that the amplitudes of  inhomogenous solutions are ``balanced'' along $x$ direction at all chemical potentials, which means the mean value of $\overline{\psi^{(2)}(x)}$ is zero at all temperatures and, thus, non-superconducting. Meanwhile, our model can not only give qualitatively the same results but, more importantly, can supplement physical process from initially homogenous and superconducting states to $T_\textrm{v}$ where the translational symmetry breaking is so intense that $\overline{\psi^{(2)}(x)}=0$, as shown in Fig. \ref{evolution}. In this figure, $\langle \ocal_2 \rangle(x)$ of first excited state with a fixed $L$ developing from $\mu_\textrm{d}$ to $\mu_\textrm{v}$ is presented, where chemical potentials from small to large are marked by black, orange, purple and red lines. At $\mu=\mu_{\textrm{d}}=15.62$ where the translational symmetry starts to break, the mean value of $\overline{\langle \ocal_2 \rangle(x)}$ is the largest; afterwards, with $\mu$ becomes larger, $\overline{\langle \ocal_2 \rangle(x)}$ decreases, and eventually becomes zero at $\mu=\mu_{\textrm{v}}=24.05$.

%Kink-like condensates $\langle \ocal_2 \rangle (x)$ near the conformal coundar; where the left and right plots correspod to the first and second excited state, respectively. Black, red and blue lines in the left plot correspond to $L=0.7,~\mu_\textrm{v}=34.30$, $L=0.825,~\mu_\textrm{v}=24.05$ and $L=0.925,~\mu_\textrm{v}=19.05$, respectively; black, red and blue lines in the left plot correspond to $L=0.455,~\mu_\textrm{v}=41.67$, $L=0.555,~\mu_\textrm{v}=25.16$ and $L=0.615,~\mu_\textrm{v}=18.40$, respectively.

As we can immediately conclude from Fig. \ref{broken con}, the chemical potential $\mu_\textrm{d}$ where the translational symmetry breaks and the chemical potential $\mu_\textrm{v}$ where the condensate vanishes are closely related to the length $L$. We therefore give their relations in Fig. \ref{mu gap}.

\begin{figure}[h]
\includegraphics[height=.24\textheight,width=.34\textheight, angle =0]{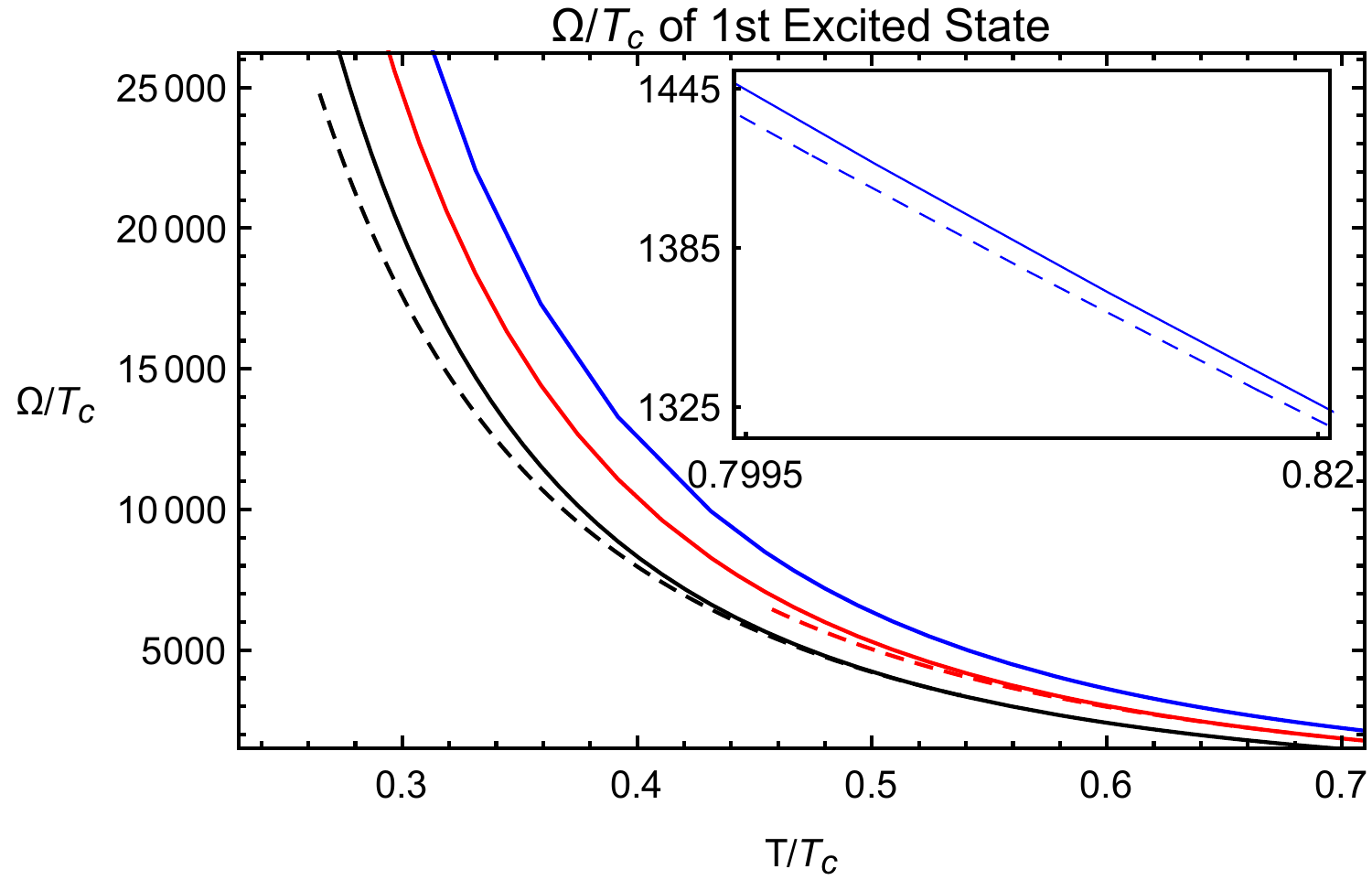}
\includegraphics[height=.24\textheight,width=.34\textheight, angle =0]{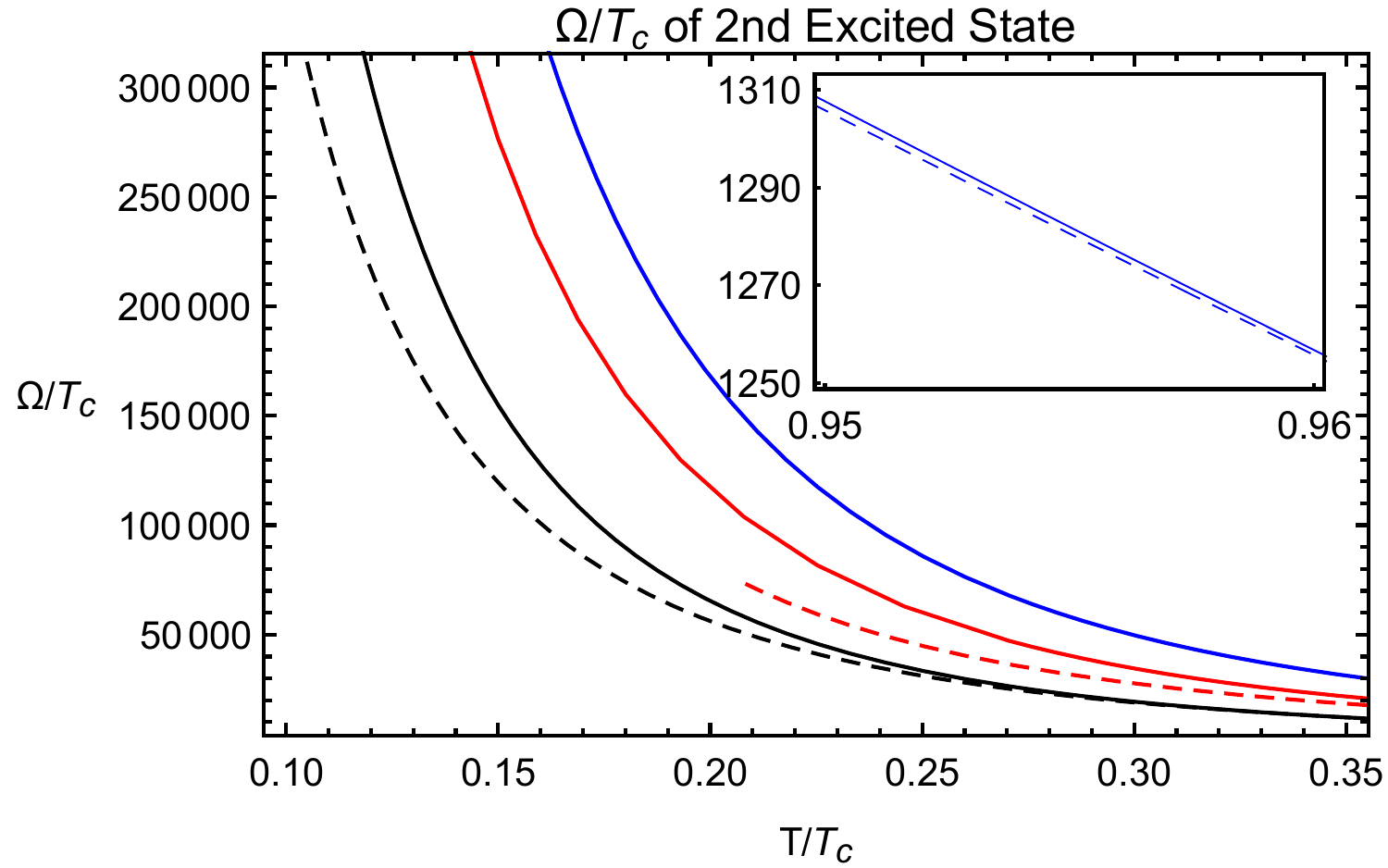}
	\caption{Free energies of the first (left panel) and second excited (right panel) states. Solid lines correspond to states that have translational symmetry, while dashed lines correspond to states where the symmetry is spontaneously  broken. Black, red and blue lines relate to different $L$ from small to large sequence. Left panel: $L=0.80~(\text{black}),~1.00~(\text{red}),~1.20~(\text{blue})$. Right panel: $L=0.255~(\text{black}),~0.455~(\text{red}),~0.655~(\text{blue})$.}\label{FE}
\end{figure}

From the figure, we can clearly see that $\mu_\textrm{d}$ is smaller than $\mu_\textrm{v}$ and, by increasing $L$, they all decrease to an identical minimum value below which there does not exist condensate; for the first excited state the minimum lattice length scale is approximately $L\approx 1.27$, the minimum value for the second excited state is approximately $L\approx 0.66$. Within the shadow regions (Red is for the first excited state, blue is for the second), the states are superconducting while the translational symmetry is spontaneously broken.

\subsection{Free energy}\label{section 3.3}
In this section, we study the free energy of the holographic superconductor in both situations with and without translational symmetry in excited states. For our model, the free energy is expressed as follow \cite{Matsumoto:2020mgi},

\begin{eqnarray}
	\frac{\Omega}{\textrm{Vol}}=\frac{1}{L}\int_{0}^{L} dx \left[\frac{1}{2}\left[\mu\rho(x)\right]+\int_{0}^{1} dz\left[\frac{\phi(x)^2}{f}\psi(x)^2\right]  \right]\label{free}.
\end{eqnarray}
Here, the Vol$=\int dzdxdy=L\int dx$. We will compare the thermal stability that has the symmetry to those that has not.

In Fig. \ref{FE}, we plot $\Omega/T_\textrm{c}$ as a function of $T/T_\textrm{c}$ for the the first and second excited states with different $L$. Where the solid lines are excited states that are translational invarient, while dash lines correspond to those that are not. The black, red and blue lines in the two subfigures correspond to different length scale $L$ in a small to large sequence, where more details of blue lines are shown in insets. In the figure, we only give $\Omega/T_\textrm{c}$ corresponding to broken translational symmetry from $T_\textrm{d}/T_\textrm{c}$ where the symmetry of matter fields starts to break to $T_\textrm{v}/T_\textrm{c}$ where the condensate goes to zero. Within this region, the model is still superconducting.

From the figure, we can see that all these curves raise from high temperature to low temperature. Besides, for a fixed temperature, free energies of both situations also raise as $L$ increases. This simple fact can be immediately found by looking at integration (\ref{free}); when one increases $L$ and thus increases the integral domain, larger free energy value is inevitable. 

However, a surprise is that, between $T_\textrm{d}/T_\textrm{c}$ and $T_\textrm{v}/T_\textrm{c}$ where the symmetry of matter fields are broken but still superconducting, the free energies shown by dashed lines are all lower than solid lines that correspond to perfect translational symmetry. This indicates that the excited states with broken translational symmetry are more thermodynamically stable than they were studied with the symmetry.

\section{Conclusions}\label{section 4}
In our work, we extended HHH model to an ansatz that matter fields rely on $x$ coordinate except the holographic coordinate and investigated the mechanism of translational symmetry breaking of excited states. In the setup, we did not assign any periodicity as source for any ingredient. Besides, no special boundary conditions were imposed on $x$ direction as well, and thus the translational symmetry was broken spontaneously. For a fixed value of $L$, the translational symmetry of the matter fields solutions was broken spontaneously at a temperature $T_\textrm{d}\textless T_\textrm{c}$, resulting condensate to decrease; after sufficiently cooling the holographic superconductor to a temperature $T_\textrm{v}$, the condensate modeled by scalar field would vanish. We also found that, in the model, $T_\textrm{d}$ and $T_\textrm{v}$, all lower than $T_\textrm{c}$, would increase with the the increase of $L$. This indicated that the length $L$ had a limit, as we found the maximum $L$ for the first excited state was approximately $1.27$, while the maximum $L$ for the second excited state was about $0.66$.

However, we have to mention that our method of introducing translational symmetry breaking was based on excited states, where we have not found solutions in ground state with such broken symmetry. Even though, the excited states might represent new bound states of interactions between quasi-particles excited above the ground state, as studied in \cite{Li:2020omw}, were found to be meta-stable; given considerable time, these excited states would evolve back to the ground state. Therefore, before recovering real experiments or finding new phenomenon via excited states, it is necessary to find solutions in these states that can be stable. In this work, we have surprisingly found the model in excited states with broken symmetry, which was still superconducting, could be more stable than those with perfect translational symmetry. Although, they were still far from being as stable as ground state.

Based on the above findings, there can be many expansions in future studies. One of which is to build new actions that bear self-interaction terms of scalar field to find solutions with translational symmetry breaking in ground state. Moreover, as various orders were found in real high temperature superconductor \cite{Li:2014wca}, the property of our model of condensates in excited states appearing at $T_\textrm{c}$ while vanishing at $T_\textrm{v}$ due to spontaneously translational symmetry breaking can be generalized to competition of multi order parameters model; where under this situation, a much more stable mode of excited state and ground state, or between excited states coexisting, is likely to be found.

\section*{Acknowledgements}
This work is supported by National Key Research and Development Program of China (Grant No. 2020YFC2201503) and  the National Natural Science Foundation of China (Grant No.~12047501). Parts of computations were performed on the shared memory system at institute of computational physics and complex systems in Lanzhou university.

\end{document}